\documentstyle[12pt,eclepsf,a4]{article}

\def\baselinestretch{1.2}
\global\arraycolsep=1pt
\def\beq{\begin{equation}}
\def\eeq{\end{equation}}
\def\beqa{\begin{eqnarray}}
\def\eeqa{\end{eqnarray}}
\def\nm{\nonumber}

\def\l{\Lambda}

\def\F{{\cal{F}}}
\def\z{\zeta}
\def\lab{\label}
\def\Del{\Delta}
\def\wDel{\widetilde{\Delta}}

\begin{document}

\begin{titlepage}
\thispagestyle{plain}
\pagenumbering{arabic}
%\null
\vspace*{-3.0cm}

\hspace{11.0cm}
hep-th/9801036
\vspace{1.5cm}

\begin{center}
{\Large \bf 
Picard-Fuchs Equation 
and Prepotential }
\end{center}
\vspace{-7.0mm}
\begin{center}
{\Large \bf of Five Dimensional SUSY 
Gauge Theory}
\end{center}
\vspace{-7.0mm}
\begin{center}
{\Large \bf Compactified on a Circle }
\end{center}

\lineskip .80em
\vskip 2em
\normalsize
\begin{center}
{\large Hiroaki Kanno}
\vskip 1.0em 
{\em Department of Mathematics\\ 
Hiroshima University \\ 
Higashi-Hiroshima 739, Japan}
\vskip 1.0em 
{and }
\vskip 1.0em
{\large Y\H uji Ohta}
%\footnote{Corresponding author, e-mail: ota@kurims.kyoto-u.ac.jp}
\vskip 1.0em
{\em Research Institute for Mathematical Sciences\\
Kyoto University \\
Sakyoku, Kyoto 606, Japan}
\vskip 1.0em
\end{center}
\vskip0.5em

\begin{abstract}
Five dimensional supersymmetric gauge theory compactified on a circle
defines an effective $N=2$ supersymmetric theory for massless fields
in four dimensions. Based on the relativistic Toda chain Hamiltonian
proposed by Nekrasov, we derive the Picard-Fuchs equation on the moduli
space of the Coulomb branch of $SU(2)$ gauge theory.
Our Picard-Fuchs equation agrees with those from other approaches;
the spectral curve of XXZ spin chain and supersymmetric cycle
in compactified $M$ theory.  By making use of a relation to
the Picard-Fuchs equation of $SU(2)$ Seiberg-Witten theory,
we obtain the prepotential and the effective coupling constant
that incorporate both a perturbative effect of Kaluza-Klein modes
and a non-perturbative one of four dimensional instantons.
In the weak coupling regime we check that the prepotential 
exhibits a consistent behavior in large and small radius limits
of the circle.
\vspace{-3.0mm}
\begin{flushleft}
\end{flushleft}

\end{abstract}

\end{titlepage}

%%%%%%%%%%%%%%%%%%%%%%%%%%%%%%%%
\def\baselinestretch{1.2}
%%%%%%%%%%%%%%%%%%%%%%%%%%%%%%%

\begin{center}
\section{Introduction}
\end{center}

\renewcommand{\theequation}{1.\arabic{equation}}\setcounter{equation}{0}

The seminal work of Seiberg-Witten \cite{SW1,SW2} has 
produced a striking progress
in our understanding of non-perturbative dynamics of 
$N=2$ supersymmetric (SUSY) gauge theory in four dimensions.
The holomorphy and the electro-magnetic duality give constraints
on the geometry of the moduli space of the Coulomb branch.
Complex curve of hyperelliptic type and
a specific differential, the Seiberg-Witten form,
on it provide natural tools for this moduli geometry called rigid special
geometry \cite{CDF}.  
The deep meaning of these objects, which appeared auxiliary, 
has been made clear from the viewpoint
of Calabi-Yau compactification of type II strings \cite{KLMVW} and also 
from $M$ theory \cite{Wit1}.

Recent developments in dualities of superstrings and $M$ theory have
made it more practical and important to consider higher dimensional
SUSY gauge theories that are perturbatively non-renormalizable.
Among them in this article we will bring into focus the five dimensional (5D)
theory compactified on a circle \cite{Nek}. There are several reasons why
this theory is interesting. 
For example, it is a SUSY gauge theory directly related 
to the Calabi-Yau compactification
of $M$ theory and its type IIA limit. It also gives us various tests on
the idea of the low energy effective action of SUSY gauge theories.
Moreover, 5D gauge theory is linked to topological
field theory and current algebra in four dimensions \cite{Nek,LMNS,BLN}.

There is an intriguing connection of the Seiberg-Witten theory to the
theory of integrable systems.
The hyperelliptic curve of pure gauge theory can be identified as
the spectral curve of the periodic Toda system \cite{MW,NT}. 
Furthermore, the Seiberg-Witten
form is given by the canonical one-form \lq$pdq$\rq\ of the symplectic dynamics.
Finally the prepotential is related to the tau function of the integrable
system.
The compactified 5D SUSY gauge theory seems to be 
accommodated to this idea.  Nekrasov proposed a relativistic
generalization of the periodic Toda system as an underlying integrable
system for 5D theory \cite{Nek}. More generally the periodic spin chain is
to be 
associated with five and six dimensional theories compactified to four
dimensions \cite{GMMM,GGM,MM}.

It has been argued that the 5D SUSY gauge theory has
no instanton corrections \cite{Sei,IMS}. 
However, it is no longer true, if we compactify
the theory to an effective four dimensional one.
In addition to the usual four dimensional one-loop corrections we expect
one-loop contributions from an infinite tower of the Kaluza-Klein modes.
There should be also instanton corrections. Both effects are supposed
to be encoded in the complex curve for the periods and the prepotential.

In this paper we will work out the Picard-Fuchs equation
for the periods of 5D SUSY $SU(2)$ gauge theory. 
Though the curve itself looks quite similar to the one for four dimensional
theory, the identification of moduli parameter is different. 
The $S^1$ compactification introduces the radius $R$ of the circle. 
This is an additional dimensionful parameter to the dynamical scale 
parameter $\Lambda$ and the combination $\zeta= \Lambda R$ is dimensionless.
Due to these changes the analysis of the Picard-Fuchs equation
becomes more complicated in five dimensional theory. 
The appearance of dimensionless parameter makes the solution more involved
and the limit $R \rightarrow 0$ more subtle.

In section 2, the Picard-Fuchs equation is derived from the spectral
curve of the relativistic Toda system. We check that the perturbative part 
of the effective coupling derived from our Picard-Fuchs equation 
incorporates the contribution from the Kaluza-Klein modes as expected. 
We also show that the Picard-Fuchs equations obtained from 
curves in other approaches; the XXZ spin chain and M theory viewpoint, 
are the same as ours. 
The periods are obtained by solving the Picard-Fuchs equation in section 3. 
We express both vector multiplet $A$ and its dual $A_D$ as power series 
in the gauge invariant moduli parameter $U$ at $U \rightarrow \infty$. 
In section 4 we establish non-linear differential equations for
the moduli parameter $U$ and find the prepotential ${\cal F}_5$ 
in the weak coupling.
We try to identify the instanton contribution in the non-perturbative 
part of the prepotential by comparing it with the prepotential 
$\F_4$ in four dimensions.
The final section is devoted to miscellaneous topics. We briefly report 
the prepotential in strong coupling region and give a preliminary result
on the $SU(3)$ case. Many technical informations are collected in appendices. 

%%%%%%%%%%%%%%%%%%%%%%%%%%%%%%%%%%%%%%%%%%%%%

\begin{center}
\section{Curve and Picard-Fuchs equation}
\end{center}

\renewcommand{\theequation}{2.\arabic{equation}}\setcounter{equation}{0}

\begin{center}
\subsection{Elliptic curve}
\end{center}

Nekrasov's proposal \cite{Nek} for the description of the moduli space 
of supersymmetric Yang-Mills gauge theory in five dimensions \cite{Sei} 
with one compactified direction was to use an elliptic curve arising 
from the Hamiltonian of the relativistic Toda chain. For $SU(2)$ 
the curve is given by  
        \beq
        y^2 =(x^2 -\Lambda^4 )\left[x-\frac{1}{2}\left(R^2 U^2 -\frac{1}{R^2}
        \right)\right]
        \lab{eli}
        ,\eeq
where $R$ is the radius of a circle and 
$U$ is the moduli parameter of the five dimensional theory. 
In the semi-classical limit $U$ has an asymptotic behavior $U \sim
R^{-2} \cosh 2RA$ \cite{Nek}, 
where $A$ is a vector multiplet whose scalar component 
has expectation value on the Coulomb branch. The 
prepotential ${\cal F}_5 (A)$ that 
characterizes the low energy effective action is locally a function of $A$.
This curve is the same as the usual $SU(2)$ Seiberg-Witten elliptic 
curve \cite{SW1} (see also appendix A),
if the moduli $u$ of the four dimensional theory is related with $U$ by 
        \beq
        u=\frac{1}{2}\left(R^2 U^2 -\frac{1}{R^2}\right)
        \lab{uU}
        .\eeq
Both the moduli parameters $u$ and $U$ have mass dimension two.

The elliptic curve (\ref{eli}) has four branching points 
        \beq
        x_1 =-\Lambda^2 ,\ x_2 =\Lambda^2 ,\ x_3 =-\frac{1}{2R^2}+
        \frac{R^2 U^2}{2},\ x_4 =\infty
        .\eeq
With the identification (\ref{uU}) these branching points correspond to
those of the Seiberg-Witten curve. Thus 
the $\alpha$- and $\beta$-cycles are taken as loops going around
counterclockwise 
as follows      
        \beq
        \alpha\mbox{-cycle} :\ x_1 \longrightarrow x_2 ,\quad
        \beta\mbox{-cycle} : \ x_2  \longrightarrow x_3 
        .\eeq
Then the intersection number is $\alpha \cap \beta =+1$. 

The periods are defined by the integral over these cycles 
of the holomorphic one form $dx/y$ on the curve.
According to Seiberg and Witten \cite{SW1,SW2} we should 
identify the periods as
        \beq
        \frac{dA}{dU}:=\frac{\sqrt{2}}{8\pi}\oint_{\alpha}\frac{dx}{y},\quad
        \frac{dA_D}{dU}:=\frac{\sqrt{2}}{8\pi}\oint_{\beta}\frac{dx}{y}
        ,\lab{peri}
        \eeq
where $A_D$ is a dual vector multiplet.
We have chosen the numerical factor so that these periods may
be identified with those of the four dimensional theory 
after the replacement (\ref{uU}). Namely, we observe the following basic
relation
to the pure $SU(2)$ Seiberg-Witten theory 
        \beq
        \frac{dA}{dU}=\frac{da}{du},\quad \frac{dA_D}{dU}=\frac{da_D}{du}
        \lab{Aa}
        .\eeq
Though $A$ and $A_D$ are obtained in principle from integration over $U$ of
(\ref{peri}) or 
contour integral of $\lambda_{SWN}$ to be introduced right now, 
(\ref{Aa}) is useful for later discussion.

The five dimensional analogue of the
Seiberg-Witten one-form can be obtained from integration of $dx/y$ over $U$. 
Up to a total derivative term, this is given by
        \beq
        \lambda_{SWN}=\frac{i}{4\pi R}
        \frac{\log[R^2 U+\sqrt{-1+R^4 U^2 -2 R^2 x}]}
        {\sqrt{x^2 -\l^4}}dx
        ,\eeq
which we call the Seiberg-Witten-Nekrasov one-form. 
It is also a characteristic one-form in the Hamiltonian dynamics
of the relativistic Toda system \cite{Nek}.
We can easily check that
        \beq
        \frac{d\lambda_{SWN}}{dU}=\frac{\sqrt{2}}{8\pi}
        \frac{dx}{\sqrt{(x^2 -\l^4 )[x-(R^2 U^2 -1/R^2 )/2]}}
        .\eeq
We also note that
        \beq
        \frac{d\lambda_{SWN}(U)}{dU}=\frac{d\lambda_{SW}(u)}{du}
        .\eeq
if we use (\ref{uU}).

\begin{center}
\subsection{The Picard-Fuchs equation}
\end{center}

The most useful tool to study the moduli space of the Coulomb branch
would be the Picard-Fuchs equation for the periods. 
In four dimensions, this method was widely 
used for many cases, e.g., for the $SU(2), SU(3), E_6,G_2$ gauge 
theories \cite{KLT,Ito,GSA,IY} and 
some algorithms to get a closed form of Picard-Fuchs equation were 
developed in \cite{Ali1,IMNS}. This method is also available 
for five dimensional case. The Picard-Fuchs equation to be derived below is 
the first example for five dimensional SUSY gauge theory. 

We can easily find that the period integrals in (\ref{peri}) satisfy the
following 
Picard-Fuchs equation
        \beq
        \frac{d^3 \Pi}{dU^3}+\left[\frac{4R^4 U}{\Delta (U)}
        (1-R^4 U^2)-\frac{1}{U}
        \right]\frac{d^2 \Pi}{dU^2}-\frac{R^8 U^2 }{\Delta (U)}
        \frac{d\Pi}{dU}=0
        ,\lab{pf}
        \eeq
where $\Pi =\oint_{\gamma}\lambda_{SWN}$ for a one-cycle $\gamma$.
        \beq
        \Delta (U)=-1+4\zeta^4 +2R^4 U^2 -R^8 U^4
        \eeq
is the discriminant of (\ref{eli}) and hereafter we will often use a
dimensionless
combination $\z =\l R$. 
Note that (\ref{pf}) is invariant under the sign 
reflection $U\rightarrow -U$ \cite{Nek}. By (\ref{uU}) 
$\Del (U)$ can be transformed into  
        \beq
        \Del (U(u))=4R^4 (\l^4 -u^2 )
        ,\lab{214}
        \eeq 
which is just the discriminant of the Seiberg-Witten curve. Similarly
we can see a relation between (\ref{pf}) and the Picard-Fuchs equation 
in four dimensions. In fact, with the help of (\ref{Aa}) we find that by the
change of variables $(U, A) \rightarrow (u, a)$,
(\ref{pf}) is reduced to 
        \beq
        \frac{d}{du}\left[4(\Lambda^4 -u^2 )\frac{d^2 \Pi}
        {du^2}-\Pi \right]=0
        \lab{pff2}
        .\eeq
Then we can integrate once to obtain 
the $SU(2)$ Picard-Fuchs equation in four dimensions. 
We note that in terms of the moduli $U$ of five 
dimensional theory (\ref{pf}) is never 
total derivative and it cannot be reduced to the 
second order equation for $\Pi$.

We can see that the Picard-Fuchs equation (\ref{pf}) incorporates 
the contribution of the (perturbative) Kaluza-Klein modes of the $S^1$ 
compactification by examining the perturbative effective coupling 
$\tau_5^{pert}$.
The effective coupling is given by the second 
derivative of the prepotential
        \beq
        \tau_5 = \frac{\partial^2 {\cal F}_5 (A)}{ \partial A^2} = \frac{d
A_D}{dA} 
         = \frac{dA_D/dU}{dA/dU}
        .\eeq
Since this is the ratio of two independent solutions of (\ref{pf}), which
is second 
order in $d\Pi/ dU$, it should satisfy the Schwarzian differential equation
(see appendix C),
        \beq
        \{ \tau_{5},U\}=-\frac{1}{2}P^2 -\frac{dP}{dU}+2Q
        ,\eeq
where $\{ *,*\}$ is called Schwarzian derivative and 
        \beq
        P=\frac{4R^4 U}{\Delta (U)}(1-R^4 U^2 )-\frac{1}{U},
        \quad Q=-\frac{R^8 U^2 }{\Delta (U)}
        ,\eeq
are the coefficients of (\ref{pf}).  To estimate the perturbative part, which is
independent of the dynamical scale $\Lambda$, 
we will formally take the limit $\Lambda \rightarrow 0$, which
makes the discriminant degenerate,
        \beq
        \Delta(U)= -(1-R^4 U^2)^2
        .\eeq
Then the differential equation for the perturbative effective coupling
$\tau_5^{pert}$ is 
        \beq
        \{ \tau_{5}^{pert},U\}= \frac{2 R^8 U^2} {(R^4 U^2 -1)^2} - 
         \frac{3}{2}\frac{1}{U^2}
        .\eeq
It is easy to see that
        \beq
        \tau_{5}^{pert}= \log (R^4 U^2 -1)
        ,\eeq
is a solution. The asymptotic behavior $UR^2 \sim \cosh 2RA$ \cite{Nek}, 
which is valid in semi-classical region, implies
        \beq
        \tau_{5}^{pert} (A) \sim \log (\sinh^2 2RA) 
        .\eeq

By the infinite product expansion of $\sinh$ we recognize the contribution 
of an infinite tower of the Kaluza-Klein excitations;
        \beq
        \tau_5^{KK} 
        =\sum_{n=1}^{\infty}\log 
        \left[A^2 +\left(\frac{n\pi}{2R}\right)^2\right]       
        ,\eeq
to the perturbative effective coupling $\tau_5^{pert}$.
We have discarded an infinite term in this manipulation.
Such a divergence could appear, since we formally turned off
the scale parameter $\Lambda$ that plays a role of 
a cut off in perturbative calculation.

%%%%%%%%%%%%%%%%%%%%%%%%%%%%%%%%%%%%%%%%%%%%%%%%%%%%%%%%

\begin{center}
\subsection{Curves from the integrable system and $M$ theory}
\end{center}

We have derived the Picard-Fuchs equation (\ref{pf}) 
from the elliptic curve (\ref{eli}) and its periods, but 
the same differential equation can be obtained from the spectral 
curve and associated one-form of 
integrable system; XXZ spin chain which includes a relativistic Toda chain 
as a particular limit \cite{GGM,MM,Mar1,Mar2,MMM}. 
Moreover, the curves of the same type have been shown to be 
obtained from the SUSY cycles in compactified $M$ theory \cite{BISTY}. 
For integrable (periodic) spin chain with $N$-sites the spectral curve is 
defined by
        \beq
        {\hbox {det}}~(T_N(\lambda) - z) 
        = 0,
        \eeq
where $T_N(\lambda)$ is the $2\times 2$ transfer (monodromy)
matrix. Thus, we obtain a
complex curve
        \beq
        z^2 - 2P_N(\lambda) z + Q_{2N}(\lambda) =0
        ,
        \eeq
with $2P_N(\lambda)= {\hbox {tr}}~T_N(\lambda)$ and 
$Q_{2N}= {\hbox {det}}~T_N(\lambda)$. 
According to \cite{GGM,MM}, pure gauge theory should be regarded as a
degenerating limit of the model with massive matter 
described by the XXZ spin chain. In our notation and normalization
the spectral curve for $SU(N_c)$ pure gauge theory is 
        \beq
        z^2 +2zx^{N_c /2}
        \prod_{i=1}^{N_c}\frac{1}{R'}\sinh R' (v-A_i ) 
        +{\l}^{2N_c} =0 
        ,\eeq
where $R' =R/\sqrt{2}$ and $x=e^{2R'v}$. Defining 
        \beq
        z+\frac{\l^{2N_c}}{z} =-\frac{2}{(2R')^{N_c}}
        P_{N_c},  \quad z-\frac{\l^{2N_c}}{z} =-\frac{2}{(2R')^{N_c}}
        y,
        \eeq
with
        \beq
        P_{N_c}=(x^{N_c}+\cdots +(-1)^{N_c})
        ,\eeq
we obtain a hyperelliptic curve;
        \beq
        y^2 =P_{N_c}^2 -(\sqrt{2}\z)^{2N_c}
        .\eeq
In the case of $N_c =2$, this is elliptic;
        \beq
        y^2 =(x^2 +s x+1)^2 -4\z^4
        , \lab{su2c}
        \eeq
where $s=-2\cosh 2R' A$. 
For the $SU(N_c)$ theory coupled 
with massless $N_f$ matter hypermultiplets, the curve takes the form 
        \beq
        z^2 +2zx^{N_c /2}
        \prod_{i=1}^{N_c}\frac{1}{R'}\sinh R' (v-A_i ) 
        +{\l}^{2N_c -N_f }x^{N_f /2}
        \frac{\sinh^{N_f} R' v}{R^{' N_f}} =0 
        .\eeq
As we have remarked these curves also arise as the SUSY cycles 
in $M$ theory compactification.

For a period integral $d\widetilde{\Pi}/ds=\oint_{\gamma}dx/y$ along 
a 1-cycle $\gamma$, the 
Picard-Fuchs equation for (\ref{su2c}) is easily deduced to be 
        \beq
        \frac{d^3
        \widetilde{\Pi}}{ds^3}+\left[\frac{1}{\wDel(s)}
        (-16s+4s^3)-\frac{1}{s}
        \right]\frac{d^2 \widetilde{\Pi}}{ds^2}+\frac{s^2}{\wDel (s) }
        \frac{d\widetilde{\Pi}}{ds}=0
        ,\lab{pppp}
        \eeq    
where 
        \beq
        \wDel (s)=16-64\z^4 -8s^2 +s^4
        .\eeq
(\ref{pppp}) is identical to (\ref{pf}) under the identification 
        \beq
        s=-2R^2 U
        .\lab{rela}
        \eeq
Therefore, both of the complex curves based on the spin chain model and on 
$M$ theory give the same Picard-Fuchs equation as the one we have
derived in sect 2.2.

We note that (\ref{su2c}) can be converted to the hyperelliptic curve in 
four dimensions 
        \beq
        \widetilde{y}^2 =(\widetilde{x}^2 -u)^2 -\l^4
        \eeq
by the transformation 
        \beq
        \sqrt{2}R\widetilde{x}=x+\frac{s}{2},\ y=2R^2 \widetilde{y},\ 
        2R^2 u=\frac{s^2}{4}-1  
        ,\lab{229}
        \eeq
where the third equation with (\ref{rela}) implies (\ref{uU}). 
However, this isomorphism breaks down 
when hypermultiplets are introduced. 
For example, let us consider the $SU(2)$ theory with a 
massless hypermultiplet (the flavor index $N_f =1$). Then 
the hyperelliptic curve derived in the $M$ theory approach 
can be written in the form \cite{BISTY}
        \beq
        y^2 =(x^2 +s x+1)^2 -(\l R)^3 (x-1)
        ,\lab{su2m}
        \eeq
where we are not careful with the numerical factor for $(\l R)^3$ 
because it can be absorbed by rescaling of variables. However, 
this curve cannot be reduced to the 
corresponding massless $SU(2)$ curve in four dimensions even if 
we apply a transformation like (\ref{229}). In this sense, 
the isomorphism via the simple transformation of variables 
between the curves of the five and four dimensional theories is 
valid only for pure gauge theories. For the models including 
hypermultiplets, we have to compare physics from 
spectral curves of integrable systems and hyperelliptic curves
of $M$ theory further. This would tell us which curve we should use.

Finally, we comment a little more on (\ref{su2m}). If (\ref{su2m}) 
is rewritten as 
        \beq
        y^2 =(\widetilde{x}^2 -u)^2 -\z^3 \left(
        \widetilde{x}-\frac{s}{2}-1\right)
        \lab{su2mm}
        \eeq
by the transformation 
        \beq
        \widetilde{x}=x+\frac{s}{2},\ u=\frac{s^2}{4}-1
        ,\eeq
(\ref{su2mm}) is reminiscent of the curve of massive $SU(2)$ $N_f =1$ 
theory in four dimensions \cite{HO} with a ``mass'' of $-(1+s/2)$ 
and the ``QCD scale parameter'' $\z$. 
Of course, since the mass of the hypermultiplet and the moduli are 
different object, we cannot actually interpret (\ref{su2mm}) 
as the massive curve in four dimensions. However, 
the reader might recall that there was a similar phenomenon 
in the $G_2$ gauge theory in four dimensions \cite{AAM}. 
There, the $G_2$ curve looked like the $SU(3)$ curve coupled with 
two massive hypermultiplets with equal mass. We do not know 
what it means, but it may imply that the moduli spaces of such different 
theories can be connected in a sense.

%%%%%%%%%%%%%%%%%%%%%%%%%%%%%%%%%%%%%%%%%%%%%%%%%%%%%%%%%%%%%%%%%%%5

\begin{center}
\section{Periods}
\end{center}

\renewcommand{\theequation}{3.\arabic{equation}}\setcounter{equation}{0}

Since the Picard-Fuchs equation (\ref{pf}) is 
essentially a second order differential equation, if we solve it 
for $d\Pi/dU$,  (\ref{pf}) can be easily solved by the Frobenius' method. 
In order to compare the results of the five dimensional theory with 
those of the corresponding four dimensional theory, 
we want to solve it in the weak coupling limit. 
As is easy to see, from the relation (\ref{uU}), the 
weak coupling regime of $u$, i.e., $u=\infty$, corresponds to $U=\infty$. 
Therefore, 
we have the following two solutions around $U=\infty$ ($z=1/U$):
        \beqa
        \rho_1 (z)&=&z\sum_{i=0}^{\infty}a_i z^i ,\nm\\
        \rho_2 (z)&=&\rho_1 (z)\log z+z\sum_{i=1}^{\infty}
        b_i z^i 
        ,\eeqa
where the first several expansion coefficients are 
        \beqa
        & &a_0 =1,\nm \\
        & &a_2 =\frac{1}{2R^4},\nm\\
        & &a_4 =\frac{3(1+2\zeta^4 )}{8R^8},\nm\\ 
        & &a_6 =\frac{5(1+6\zeta^4 )}{16R^{12}},\nm\\ 
        & &a_8 =\frac{35(1+12\zeta^4 +6\zeta^8 )}{128R^{16}},\nm\\ 
        & &a_{10}=\frac{63(1+20\z^4 +30\z^8 )}{256R^{20}},\nm\\
        & &a_{12}=\frac{231(1+30\z^4 +90\z^8 +20\z^{12})}{1024R^{24}}   
        ,\nm\\
        & &a_{14}=\frac{429(1+42\z^4 +210\z^8 +140\z^{12})}{2048R^{28}}
        \eeqa
and 
        \beqa
        & &b_2 =\frac{1}{2R^4},\nm\\
        & &b_4 =\frac{4+5 \zeta^4}{8R^8},\nm\\
        & &b_6 =\frac{23+93\zeta^4}{48R^{12}},\nm\\ 
        & &b_8 =\frac{352+2964\zeta^4 +1167\zeta^8}{768R^{16}},\nm\\
        & &b_{10}=\frac{1126+16220\z^4 +19605\z^8}{2560R^{20}},\nm\\
        & &b_{12}=\frac{13016+286530\z^4+703665\z^8 +133270\z^{12}}{
        30720R^{24}},\nm\\
        & &b_{14}=\frac{176138+5505906\z^4 +22799805\z^8 +
        13097770\z^{12}}{430080R^{28}}
        .\lab{coe}
        \eeqa   
We note that $a_{2n+1}=b_{2n+1}=0$ due to the invariance 
under $U\rightarrow -U$.

The period integrals (\ref{peri}) are given by some linear combinations of
the fundamental solutions $\rho_1$ and $\rho_2$. 
Lower order expansion fixes the combination to be
        \beq
        \frac{dA}{dU}=\frac{1}{2R}\rho_1 ,
        \quad  \frac{dA_D}{dU}=i\frac{2}{\pi R}\log\left(\frac{2R}{\l}\right)
        \rho_1 - i\frac{2}{\pi R}
        \rho_2
        \lab{inte}
        .\eeq
We must further integrate by $U$ in order to obtain $A$ and $A_D$. 
First,  let us consider $A$. Integration over $U$ gives 
        \beqa
        A&=&\mbox{integration const.}+
        \int \frac{\rho_1}{2R} dU\nm\\
        &=&\frac{1}{2R}\log 2 R^2 +\frac{1}{2R}\log U 
        -\frac{1}{8R^5U^2}-\frac{1}{U^4}
        \left(\frac{3}{64R^9}+\frac{3\l^4}{32R^5}\right)-\frac{1}{U^6}
        \left(\frac{5}{192R^{13}}+\frac{5\l^4}{32R^9}\right)\nm\\
        & &-\frac{1}{U^8}\left(\frac{35}{2048R^{17}}+\frac{105\l^4}{512R^{13}}+
        \frac{105\l^8}{1024R^9}\right)-\cdots
        ,\lab{tte}
        \eeqa
where we have identified the integration constant with 
$(\log 2R^2)/(2R)$, because the direct calculation of $A$ reads
        \beqa
        A&=&\oint_{\alpha}\lambda_{SWN}\nm\\
        &=&\frac{i}{2\pi R}\int_{-\l^2}^{\l^2}
        \frac{\log[R^2 U+\sqrt{-1+R^4 U^2 -2 R^2 x}]}
        {\sqrt{x^2 -\l^4}}dx\nm\\
        &=&\frac{1}{2R}\log 2R^2 +\frac{1}{2R}\log U -\frac{1}{8R^5 U^2}
        -\cdots
        .\eeqa
It is easy to find that $\l$-independent terms in (\ref{tte}) are
originated from 
the expansion of $(1/2R)\log(R^2 U+\sqrt{R^4 U^2 -1})$ for large $U$. 
Therefore, it follows that  
        \beq
        A=\frac{1}{2R}\log(R^2 U+\sqrt{R^4 U^2 -1})
        -\frac{3\l^4}{32R^5 U^4}
        -\frac{5\l^4}{32R^9 U^6}-\frac{1}{U^8}
        \left(\frac{105\l^4}{512R^{13}}+
        \frac{105\l^8}{1024R^9}\right)-\cdots 
        .\lab{315}
        \eeq
We expect that the remaining $\l$-dependent terms may 
be deduced from expansion of 
some simple functions, but we could not find such expressions.

Similarly, we can find the asymptotic behavior of $dA_D /dU$
        \beq
        \frac{dA_D}{dU}\sim i\frac{R}{\pi \sqrt{R^4 U^2 -1}}\log 
        \frac{4(R^4 U^2 -1)}{\z^2}
        .\eeq
Integration over $U$ gives the leading part of $A_D$ and it is 
represented by using 
di-logarithm (see appendix D), but the explicit form is not so simple 
and elegant, so we do not 
give it here. 
We do not need to include a integration 
constant in the expansion of $A_D$, because 
it has no effect on prepotential or monodromy. Note that the 
integration constant 
term corresponds to a linear term in the prepotential.

Since the periods are determined, their monodromy 
can be now obtained by leading terms
        \beqa
        A&=&\frac{1}{2R}\log 2R^2 +\frac{1}{2R}
        \log U -\cdots ,\nm\\
        A_D &=&i\frac{4}{\pi}A\log\frac{2RU}{\l}
        -i\frac{2}{\pi R}\log(2R^2)
        \log U -i\frac{1}{\pi R}\log^2 U +\cdots 
        .\eeqa
Then the monodromy $U\rightarrow e^{2\pi i}\cdot U$ around $U=\infty$ 
transforms these periods into 
        \beqa
        A&\longrightarrow & A+i\frac{\pi}{R},\nm\\
        A_D&\longrightarrow&A_D -8A-i\frac{4\pi}{R}
        +\frac{4}{R}\log \z 
        .\eeqa
Thus the monodromy matrix is given by 
        \beq
        \left(
        \begin{array}{c}
        A_D \\
        A \\
        1/R \end{array}\right) \longrightarrow \left(
        \begin{array}{ccc}
        1&-8& m\\
        0&1&i\pi \\
        0&0&1 \end{array}\right)\left(
        \begin{array}{c}
        A_D \\
        A \\
        1/R  \end{array}\right)
        ,\eeq
where $m=-4\pi i +4\log \z $.

For completeness, let us consider the strong coupling region. 
The strong coupling region in four dimensions 
is located at $u=\pm \l^2$ where 
a magnetically charged particle and a dyon become massless, 
respectively, but in this five dimensional 
theory the corresponding singularities in the moduli space 
are splitted into 
        \beq
        U_{+}^{\pm}:=\pm \frac{1}{R^2}\sqrt{1+2\z^2}, \quad 
        U_{-}^{\pm}:=\pm \frac{1}{R^2}\sqrt{1-2\z^2}
        ,\eeq
respectively (see also figure 1). 
%%%%%%%%%%%%%%%%%%%%%%%%%%%%%%%%%%%%%%%%%%%%%%%%%%%%%%%%%%%
%% The authors recommend here as the place of the figure.
%%%%%%%%%%%%%%%%%%%%%%%%%%%%%%%%%%%%%%%%%%%%%%%%%%%%%%%%%%%%%

Let us consider a singularity corresponding to $u=+\l^2$. In this case, 
we may take either $U_{+}^+$ or $U_{+}^-$, but we take the former. 
By a similar discussion to the weak coupling case, the periods in the strong 
coupling regime can be easily found. However, since the solutions to 
the Picard-Fuchs equation 
are represented by a series with lengthy expansion coefficients, we 
consider equivalent 
expressions which can be deduced from the periods in four dimensional 
theory in the strong coupling regime (recall that 
the periods are related in the sense of (\ref{Aa})). The result 
is ($\widetilde{z}=U-\sqrt{1+2\z^2}/R^2 $): 
        \beqa
        \frac{dA_D}{dU}&=&i\frac{\l}{2}\sum_{n=0}^{\infty}\widetilde{a}_n 
        (n+1)\left[\frac{(\z/\l)^2 \widetilde{z}^2 
        +2\widetilde{z}\sqrt{1+2\z^2}}{4\l^2}\right]^n ,\nm\\
        \frac{dA}{dU}&=&\frac{i}{2\pi}\frac{dA_D}{dU} 
        \left[\log \frac{(\z/\l)^2 \widetilde{z}^2 
        +2\widetilde{z}\sqrt{1+2\z^2}}{4\l^2} -1-\log 2\right]\nm\\
        & &-\frac{\l}{4\pi}\sum_{n=0}^{\infty}
        \left[
        \widetilde{a}_n +\widetilde{b}_n (n+1)\right]
        \left[ \frac{(\z/\l)^2 \widetilde{z}^2 
        +2\widetilde{z}\sqrt{1+2\z^2}}{4\l^2}\right]^n 
        ,\lab{strperi}
        \eeqa
where  
        \beqa
        \widetilde{a}_n &=&(-1)^n \frac{(1/2)_{n}^2 }{n! (2)_n},\nm\\
        \widetilde{b}_n &=&\widetilde{a}_n \left[ 2\left[\psi \left(n
        +\frac{1}{2}\right) 
        -\psi\left(\frac{1}{2}\right)
        \right] +\psi (1) -\psi (n+1 )+\psi (2) -\psi (n+2) \right]
        .\eeqa
Furthermore, $(*)_n$ is the Pochhammer's symbol 
and $\psi (*)$ is the polygamma function. 
Precisely speaking, we should expand these expressions around 
$\widetilde{z}=0$.        

Integration over $U$ provides the periods in the strong 
coupling regime and then 
the monodromy at $U=U_{+}^+$ is calculated from the asymptotic property  
        \beqa
        A_D &=&\frac{i}{2\l}(U-U_{+}^+ )+\cdots ,\nm\\
        A&=& \frac{i}{2\pi}A_D \log (U-U_{+}^+ )+\cdots 
        .\eeqa
Thus the monodromy matrix is given by
        \beq
        \left(\begin{array}{c}
        A_D \\
        A\\
        1/R \end{array}\right) \longrightarrow \left(\begin{array}{ccc}
        1&0&0\\
        -1&1&0\\
        0&0&1
        \end{array}\right)\left(\begin{array}{c}
        A_D \\
        A\\
        1/R \end{array}\right) 
        .\eeq
We note that in contrast with the weak coupling case, $1/R$ does not mix
with the monodromy 
of $A$ and $A_D$ in the strong coupling.

%%%%%%%%%%%%%%%%%%%%%%%%%%%%%%%%%%%%%%%%%%%%%%%%%%%%%%%%%

\begin{center}
\section{Prepotential in weak coupling limit}
\end{center}

\renewcommand{\theequation}{4.\arabic{equation}}\setcounter{equation}{0}

\begin{center}
\subsection{The prepotential}
\end{center}

By the basic relation of special geometry;
       \beq
       A_D = \frac{\partial {\cal F}_5}{\partial A}~, \quad
       \tau_5 = \frac{\partial^2 {\cal F}_5}{\partial A^2}~ 
        ,\eeq
the prepotential $\F_5$ in five dimensions can be obtained by 
integrating either $A_D$ or $\tau_5$ over $A$ after 
expressing them as a function of $A$. 
Hence, our first task is to find an inversion 
of the relation $A=A(U)$ obtained in the last section. 
This inversion requires some 
technical care and we take the following manner. 
Recall that in the four dimensional theory similar inversion of 
period was proceeded by solving 
a differential equation for 
the moduli \cite{Mat,BMT}. Also for the 
case at hand, we can find the required relation $U=U(A)$ 
by solving a non-linear differential equation 
for $U$ as follows.

Let 
        \beq
        u=u(a),\quad U=U (A)
        ,\lab{51}
        \eeq
then by the relation of period integrals (\ref{Aa}) we have 
        \beq
        \dot{u}= U'
        \lab{mat}
        ,\eeq
where $\ \dot{ }=d/da$ and $'=d/dA$. The second and third order 
derivatives are then given by 
        \beq
        U'' =R^2 U \ddot{u},\quad U'''=R^2 (\dot{u}\ddot{u}
        +R^2 U^2 \stackrel{...}{u})
        .\lab{53}
        \eeq
We also have the following relation;
       \beq
        \frac{dA}{dU} = (U')^{-1}~, \quad 
        \frac{d^2 A}{dU^2} = - (U')^{-3} U''~, \quad
        \frac{d^3 A}{dU^3} = 3 (U')^{-5} U^{''2}  - (U')^{-4} U'''~, 
        \lab{55}
        \eeq
Substituting (\ref{51}), (\ref{mat}) and (\ref{55}) 
into the Picard-Fuchs equation for $A$, we get 
        \beq
        U''' U' -3{U''} ^2 +\left[\frac{4R^4 U}{\Delta(U)}
        (1-R^4 U^2 )-\frac{1}{U}\right]{U'} ^2 U'' 
        +\frac{R^8 U^2 }{\Delta (U)}{U'} ^4 =0
        \lab{mat2}
        ,\eeq
which can be rewritten by using (\ref{53}) as 
        \beq
        \frac{d}{da}\left[\widetilde{\Delta}(u) \ddot{u}+a \dot{u}^3 \right]=0
        \lab{57}
        ,\eeq
provided 
        \beq
        \Del (U)=R^4 \widetilde{\Delta}(u) = 4 R^4  (\l^4 -u^2 )
        .\eeq
(\ref{57}) is nothing but the (differentiation of) Matone's 
differential equation for $u$ found in four dimensions
\cite{Mat,BMT}. 

We can now solve (\ref{mat2}), but we must take into account of 
an ``initial condition''. Note that from (\ref{315}) $A$ is asymptotically 
        \beq
        A\sim \frac{1}{2R}\log(R^2 U+\sqrt{R^4 U^2 -1})
        ,\eeq
or
        \beq
        U\sim \frac{1}{R^2}\cosh 2RA
        \lab{in}
        .\eeq
This means that we should solve (\ref{mat2}) by assuming 
        \beq
        U =\frac{1}{2R^2}\left(q+\frac{1}{q}\right)+ \frac{1}{R^2}
        \sum_{n=1}^{\infty}c_n (q)\z^{4n}, \quad (\z = \l R)
        \lab{expan}
        ,\eeq
where $q=e^{2RA}$. 
In fact, substituting (\ref{expan}) 
into (\ref{mat2}), we can obtain 
differential equations for $c_n$ by equating the coefficients of 
powers in $\z$ to zero. For higher $c_n$, such differential equations 
are very complicated in general, but we can solve it for $c_1$. The result is 
        \beq
        U=\frac{1}{R^2}\cosh 2R A+\frac{\l^4 R^2}{2}\frac{3q^2 -1}
        {q(1-q^2)^2}+O(\l^8 )
        .\lab{Uten}
        \eeq

The prepotential can be calculated by using $A_D$ (or $\tau_5$) 
and (\ref{Uten}). For example, for large $R$, the prepotential 
is given by   
        \beqa
        \F_5&=&
        i\frac{4R}{3\pi}A^3 -i\frac{2\log \zeta}{\pi}A^2 -i\frac{1}
        {4\pi R^2}Li_3 (q^{-2})\nm\\
        & &-i\frac{\l^4 R^2}{\pi}\left(\frac{7}{16q^4}+\frac{13}{18q^6}
        +\frac{63}{64q^8}+\frac{31}{25q^{10}}+\frac{215}{144q^{12}}
        +\cdots\right)-O\left(\l^8\right)
        ,\lab{prepo}
        \eeqa
where $Li_3$ is a special function called 
tri-logarithm (see \cite{Lew1,Lew2} and 
also appendix D). In this derivation, we have suppressed 
linear terms in $A$. Note that the first and the second terms 
in (\ref{prepo}) are independent of $q$. Furthermore, though in the 
large radius limit $Li_3 (q^{-2})$ vanishes exponentially, we 
can see the small radius behavior of it by considering analytic 
property of tri-logarithm (see below). Therefore, the $SU(2)$ 
prepotential in the 
five dimensional theory can be generally written as
        \beq
        \F_5=
        i\frac{4R}{3\pi}A^3 -i\frac{2\log \zeta}{\pi}A^2 -i\frac{1}
        {4\pi R^2}Li_3 (q^{-2})+\sum_{n=1}^{\infty}\l^{4n}\F_{5,n}(q,R)
        ,\lab{prepo2}
        \eeq
where $\F_{5,n}$ is a function of $q$ and $R$, in particular, we have 
        \beq
        \F_{5,1}|_{R\rightarrow \infty}=
        -i\frac{R^2}{\pi}\left(\frac{7}{16q^4}+\frac{13}{18q^6}
        +\frac{63}{64q^8}+\frac{31}{25q^{10}}+\frac{215}{144q^{12}}
        +\cdots\right)
        ,\eeq
although we could not find the exact function $\F_{5,1}$ itself (but see 
appendix E).

%%%%%%%%%%%%%%%%%%%%%%%%%%%%%%%%%%%%%%%%%%%%%%%%%%%%

\begin{center}
\subsection{Large and small radius limit}
\end{center}

Firstly, we take the large radius limit of (\ref{prepo}). 
At $R \rightarrow \infty$, divided by the length of $S^1$, 
i.e., $2\pi R$, $\log \z /R$ and $Li_3 (q^{-2})$ vanish
because of $q^{-1}\sim 0$, 
thus only the cubic term of (\ref{prepo}) survives in the
decompactification limit;
        \beq
        \left.\frac{\F_5 }{2\pi R}\right|_{R\rightarrow \infty}=
        i\frac{2}{3\pi^2}A^3 
        .\eeq
More generally the cubic term in the prepotential of (uncompactified)
five dimensional SUSY gauge theory is identified as
the intersection forms of Calabi-Yau manifolds
in the Calabi-Yau compactification of $M$ theory \cite{Sei,IMS}. 

The small radius limit is obtained as follows;        
Let us denote the leading part of $\F_5$ by 
        \beq
        \F_{5led}=i\frac{4R}{3\pi}A^3 -i\frac{2\log \zeta}{\pi}A^2 
        -i\frac{1}{4\pi R^2}Li_3 (q^{-2})
        .\eeq
Then from the asymptotic property of $Li_3$ we have 
        \beq
        \left.\F_{5led} \right|_{R\rightarrow 0}=
        i\frac{A^2}{\pi}\left[\log\left(\frac{A}{\l}\right)^2 +
        4\log 2-3\right]
        \lab{81}
        \eeq
for $R\rightarrow 0$, where we have suppressed $O(A)$ terms because they are 
irrelevant to prepotential. (\ref{81}) coincide with the leading part of
$\F_4$ if $A$ is identified with $a$ (see below). 

In the above small radius limit, we keep only the leading terms
that give rise to the one-loop prepotential in four dimensions.
The non-perturbative part of the prepotential 
in small radius limit is not clear in the naive limit,
since we have $R$-dependence in $q=e^{2RA}$.
We will discuss the instanton contributions
by a slightly different approach in the next part.

%%%%%%%%%%%%%%%%%%%%%%%%%%%%%%%%%%%%%%%%%%%%%%%%%%%%%%%%%%%%%%%%

\begin{center}
\subsection{Instanton contributions to the prepotential}
\end{center}

We want to see how instantons contribute to the 
prepotential in the five dimensional theory. However, 
we do not know an exact expression of $\F_{5,n}$ in (\ref{prepo2}). 
To proceed we will take the following way.
Recall that we have normalized 
the relation between four and five dimensional theories 
such that their effective coupling constants are equivalent by (\ref{Aa}). 
Therefore, even if $\tau_5$ is written by $A$, it is equivalent to $\tau_4$, 
so the four and five dimensional theories are equivalent in 
this sense because $\tau_5$ is defined by
        \beqa
        \tau_{5}&=&\frac{dA_D}{dU}/\frac{dA}{dU}\nm\\
        &=&\frac{da_D}{du}/\frac{da}{du}\nm\\
        &=&\tau_4 
        \lab{efc}
        ,\eeqa
where $\tau_4$ is the effective coupling constant in four dimensions. 
Moreover, $\tau_4$ as a function of $a$ is obtained 
from differentiation of $\F_4$ 
        \beqa
        \tau_4 &=&\frac{d^2 \F_4}{da^2}\nm\\
        &=&i\frac{8}{\pi}\log 2+i\frac{2}{\pi}\log \left(\frac{a}{
        \l}\right)^2 +i\frac{2}{\pi}\sum_{k=1}^{\infty}
        (2k-1)(4k-1)F_k \left(\frac{\l}{a}
        \right)^{4k}
        \lab{hati}
        .\eeqa
Therefore, if $a$ can be expressed by $A$, then twice integration over $A$ 
will generate a prepotential of the five dimensional theory. Then this
prepotential 
will include the instanton expansion coefficients $F_k$ in four dimensions. 

Based on this observation, we can calculate the prepotential as follows; 
Note that the transformation of variables between $a$ and $A$ is found by
        \beq
        \frac{da}{dA}=\frac{du}{dU}
        ={R^2 U}
        \lab{hasan}
        \eeq
with (\ref{Aa}) and (\ref{uU}). If we integrate (\ref{hasan})
after substitution of (\ref{Uten}),
$a$ is represented by $A$. Thus, we find  
        \beq
        a=\frac{1}{2R}\sinh 2RA
        -\frac{\l^4 R^3}{4q}\frac{1}{q^2 -1}-O(\l^8 )
        \lab{aex}
        ,\eeq
where the integration constant has been set to zero
by the 
following reason. Assume that the leading part of (\ref{aex}) is given by       
        \beqa
        a&\sim&c~(\mbox{= integration const.})+\frac{1}{4R}\left(q-\frac{1}{q}
        \right)\nm\\
        &=&c+\frac{1}{2R}\sinh 2RA
        .\lab{87}
        \eeqa
Then, (\ref{87}) means
        \beqa
        R^4 U^2 &=&\cosh^2 2RA\nm\\
        &=&1+4R^2 (a-c)^2
        .\eeqa
But the leading part of $u$ as a function of $a$ is known to be $2a^2$ and 
from (\ref{uU}) we conclude that $c=0$.

By expanding $\tau_4$ by $q$ after substitution of (\ref{aex}) into
(\ref{hati}), 
we obtain $\tau_4$ as a function of $A$. In view of (\ref{efc}) twice 
integration of $\tau_4 (A)$ over $A$ gives the prepotential
        \beqa
        \widehat{\F}_5 &=&i\frac{4R}{3\pi}A^3 -i\frac{2\log \zeta}
        {\pi}A^2 -i\frac{1}
        {4\pi R^2}Li_3 (q^{-2})\nm\\
        & &+i\frac{\l^4 R^2}{\pi}\left[\frac{1}{q^4}\left(-\frac{1}{16}
        +24F_1\right)+\frac{1}{q^6}\left(-\frac{1}{18}+\frac{128}{3}F_1
        \right)+\frac{1}{q^8}\left(-\frac{3}{64}+60F_1\right)\right.\nm\\
        & &+\left.\frac{1}{q^{10}}\left(-\frac{1}{25}+\frac{384}{5}F_1
        \right)+\frac{1}{q^{12}}\left(-\frac{5}{144}+\frac{280}{3}
        F_1\right)+\cdots\right]-O\left(\l^8 \right)
        \lab{prepin}
        .\eeqa
Using explicit numerical data (\ref{inexp}) of instanton 
expansion coefficients, we find that 
        \beqa
        \widehat{\F_5}&=&i\frac{4R}{3\pi}A^3 
        -i\frac{2\log \zeta}{\pi}A^2 -i\frac{1}
        {4\pi R^2}Li_3 (q^{-2})\nm\\
        & &-i\frac{\l^4 R^2}{\pi}\left(\frac{7}{16q^4}+\frac{13}{18q^6}
        +\frac{63}{64q^8}+\frac{31}{25q^{10}}+\frac{215}{144q^{12}}+\cdots
        \right)-O\left(\l^8\right) 
        .\eeqa
This prepotential $\widehat{\F}_5$ coincides with (\ref{prepo})!
Since we have calculated the prepotential by two essentially independent
ways and 
checked they are in agreement, we can conclude that the prepotential is 
correctly determined.

%%%%%%%%%%%%%%%%%%%%%%%%%%%%%%%%%%%%%%%%%%%%%%%%%%%%%%%%%%%%%%

\begin{center}
\subsection{Relation between $\F_5$ and $\F_4$}
\end{center}

We have shown how the instanton corrections arise in the prepotential in 
five dimensions by comparing with $\F_4$.
In this subsection, we give an argument 
to establish that $\F_5$ reduces to $\F_4$ (in the weak 
coupling regime), provided the radius of $S^1$ vanishes.  
        
Let us recall that $\tau_4 (a)=\tau_5 (A)$. Since $\F_5$ is a twice 
integrated quantity over $A$, it follows that
        \beqa
        \F_5 &=&\int\int\left(\tau_5 dA\right)dA\nm\\
        &=&\int\int\left(\tau_4 \frac{dA}{da}da\right)\frac{dA}{da}da
        .\eeqa
If $dA/da$ can be written by $a$, $\F_5$ will be a function of $a$.
Furthermore, 
if we take the limit $R\rightarrow 0$, we will obtain some relation between 
$\F_5$ and $\F_4$. For this purpose, however, 
the inversion of $a=a(A)$ is required. From (\ref{aex}), we observe that 
        \beq
        \frac{1}{2R}\sinh 2RA=a+Y
        ,\eeq
where 
        \beq
        Y=\frac{\l^4 R^3}{4q}\frac{1}{q^2 -1}+O(\l^8 ) 
        .\eeq   
Thus we find 
        \beq
        A=\frac{1}{2R}\sinh^{-1}(2Ra +2RY)
        .\lab{asin}
        \eeq
Solving (\ref{asin}) recursively in $A$ and expanding around $R=0$, 
we can obtain 
        \beq
        A=a+\left(-\frac{2}{3}a^3 +\frac{\l^4}{16 a}\right)R^2 
        -\frac{\l^4 }{4}R^3 +\cdots
        .\eeq
Therefore, 
        \beq
        \frac{dA}{da}=1+Z
        ,\lab{tends}
        \eeq
where we have denoted miscellaneous terms depending on $R$ by $Z$, which is
manifestly 
$Z\rightarrow 0$ as $R\rightarrow 0$. It is obvious that $A\sim a$ 
as $R\rightarrow 0$. 
        
Consequently, the prepotential will have the 
form 
        \beqa
        \F_5 &=&\int\int\left[ \tau_4 (1+Z)da\right](1+Z)da\nm\\
        &=& \F_4  +\int\int (\tau_4 da)Zda+
        \int\int (\tau_4 Zda)da
        +\int \int (\tau_4 Zda)Zda
        .\eeqa
This expression shows that
in the weak coupling regime, $\F_5$ reduces to $\F_4$ 
if the radius of $S^1$ vanishes;
        \beq
        \F_5 |_{R\rightarrow 0}=\F_4 
        .\eeq

%%%%%%%%%%%%%%%%%%%%%%%%%%%%%%%%%%%%%%%%%%%%%%%%%%%%%%%%%%%%%%%%%%%%%%%%

\begin{center}
\section{Miscellany}
\end{center}

\renewcommand{\theequation}{5.\arabic{equation}}\setcounter{equation}{0}

In the last sections the expansion in the weak coupling has 
been worked out. The calculation in the strong coupling regime is 
slightly easy in contrast with the weak coupling case. Repeating a 
similar construction as in the preceding sections by interchanging 
a role of $A$ and $A_D$, 
we obtain the dual prepotential by using the inversion of $A_D $ 
in (\ref{strperi});
        \beqa
        \widetilde{z}&=&-2i\l A_D -\frac{1}{4}\sqrt{1+2\z^2 }
        A_{D}^2 +i\frac{
        (-3+10\z^2 )}{96\l}A_{D}^3 -\frac{\sqrt{1+2\z^2}
        (-15+26\z^2)}{1536\l^2}A_{D}^4 \nm\\
        & &-i\frac{-495+420\z^2 +1028\z^4 }
        {122880\l^3}A_{D}^5 \nm\\
        & &+\frac{\sqrt{1+2\z^2}(-2835+5460\z^2 +116\z^4 )}
        {1474560\l^4}A_{D}^6 +\cdots
        .\lab{ADten}
        \eeqa
Also in this case we can check 
(\ref{ADten}) from (\ref{mat2}) by exchanging $A$ and $A_D$. 
Then the dual prepotential is found to be 
        \beq
        \F_{D_5}=A_{D}^2 \left[\frac{i}{8\pi}\log
        \left[(1+2\z^2)\left(\frac{A_D}{\l}
        \right)^2 \right] +\F_{D_{5,2}}
        \right] +\frac{i\l^2}{\pi}(1+2\z^2 )\sum_{k=3}^{\infty}\F_{D_5 ,k}
        \widetilde{A}_{D}^k 
        ,\eeq
where $\F_{D_{5,2}}=1/8 -i3/(8\pi)-i\log 2 /(4\pi)$,  
        \beq
        \widetilde{A}_{D} =i\frac{A_D}{\l \sqrt{1+2\z^2}}
        \eeq
and the first several dual ``instanton'' expansion coefficients 
are given by
        \beqa
        \F_{D_5 ,3}&=&\frac{-3+2\z^2}{96},\nm\\
        \F_{D_5 ,4}&=&\frac{-45-68\z^2 +236 \z^4}{9216},\nm\\
        \F_{D_5 ,5}&=&-\frac{-165-390\z^2 +292\z^4 +1848\z^6}{122880}
        ,\nm\\
        \F_{D_5 ,6}&=&-\frac{-42525-138600\z^2 +43592\z^4 
        +415328\z^6 +596528\z^8 }{88473600},\nm\\
        \F_{D_5 ,7}&=&\frac{-33201-137970\z^2 -25704\z^4 
        +502000\z^6 +475568\z^8 +275808\z^{10}}{165150720}
        .\eeqa
One can see that this dual prepotential 
coincides with the dual prepotential in four 
dimensions \cite{KLT,IY2} for $R\rightarrow 0$ if $A_D$ is 
identified with $a_D$.

In this paper we have studied the five dimensional 
$SU(2)$ theory, but we can consider higher rank gauge theories 
as well. For example, the $SU(3)$ curve in five dimensions 
takes the form 
        \beq
        y^2 =(x^3 +s_1 x^2 +s_2 x -1)^2 -(\l R)^6
        ,\lab{suu3c}
        \eeq
where $s_1$ and $s_2$ are moduli \cite{BISTY}. Under the identification 
        \beq
        s_2 -\frac{s_{1}^2}{3}=-R^2 u, \quad  1-\frac{2s_{1}^3}{27}
        +\frac{s_1 s_2}{3}=-R^3 v
        ,\eeq
(\ref{suu3c}) is shown to be equivalent 
to the curve of four dimensional 
theory \cite{KLT,KLYT1,KLYT2,AF,HO,AAG} given by 
        \beq
        y^2 =(x^3 -ux-v)^2 -\l^6
        ,\lab{su3c}
        \eeq
where $u$ and $v$ are moduli in four dimensions. 
Then the discriminant of (\ref{suu3c}) is given by    
        \beqa
        \Del_{SU(3)} &=&\left[27(1-\z^3 )^2  +(1-\z^3 )
        (-4s_{1}^3 +18s_1 s_2 )
         -s_{1}^2  s_{2}^2 +4s_{2}^3 \right]\nm\\
        & &\times \left[27(1+\z^3 )^2 +(1+\z^3 )
        (-4s_{1}^3 +18s_1 s_2 )
         -s_{1}^2  s_{2}^2 +4s_{2}^3 \right]
        ,\eeqa
which can be transformed into 
        \beq
        \Del_{SU(3)}=R^{12}\left[ 4u^3 -27 
        (\l^3 -v)^2 \right]\left[
        4u^3 -27 (\l^3 +v)^2 \right]
        .\eeq
This is the discriminant of (\ref{su3c}) (up to overall constant). 
We do not write down the $SU(3)$ Picard-Fuchs equations in five dimensions 
because it is too lengthy and complicated, but it would be interesting to 
further discuss and compare with the results of four dimensions.

\bigskip
One of the authors (H.K.) would like to thank N. Nekrasov for 
sending a copy of his thesis.  He also thanks Katsushi Ito and S.-K. Yang  
for continual discussions on the Seiberg-Witten theory.
\bigskip

%%%%%%%%%%%%%%%%%%%%%%%%%%%%%%%%%%%%%%%%%%%%%%%%%%%%%%%%%%%%%%%%%%%%%%

\begin{center}
\section*{Appendix A. The $SU(2)$ theory in four dimensions} 
\end{center}

\renewcommand{\theequation}{A.\arabic{equation}}\setcounter{equation}{0}

In this appendix, we briefly summarize the 
$N=2$ $SU(2)$ supersymmetric Yang-Mills theory 
in four dimensions. For details, see \cite{SW1,SW2,KLT}.

The Seiberg-Witten elliptic curve and the meromorphic 
1-form are given by 
        \beq
        y^2 =(x^2 -\l^4 )(x-u)
        \eeq
and  
        \beq
        \lambda_{SW}=\frac{\sqrt{2}}{4\pi}
        \sqrt{\frac{x-u}{x^2-\l^4}}dx
        ,\eeq
respectively. We choose the $\alpha$- and $\beta$-cycles as
        \beq
        \alpha-\mbox{cycle}: -\l^2 \longrightarrow \l^2 ,\ 
        \beta-\mbox{cycle}: \l^2 \longrightarrow u
        .\eeq
For these cycles, the periods are defined by 
        \beq
        a=\oint_{\alpha} \lambda_{SW},\ a_D =\oint_{\beta}
        \lambda_{SW} 
        .\eeq
Using the formulae for hypergeometric function (see Appendix B), we find 
that these periods can be represented by 
        \beqa
        \frac{da}{du}&=&\frac{\sqrt{2}}{4\sqrt{u+\Lambda^2}}\ _2 F_1 
        \left(\frac{1}{2},\frac{1}{2},1;\frac{2\Lambda^2}{\Lambda^2 +u}
        \right)\nm\\
        &=&\frac{1}{4}\sqrt{\frac{2}{u}}\ _2 F_1 \left(\frac{1}{4},\frac{3}{4},
        1;\frac{\Lambda^4}{u^2}\right),\nm\\
        \frac{da_D}{du}&=&\frac{i}{2\Lambda} \ _2 F_1 \left(
        \frac{1}{2},\frac{1}{2},1;\frac{\Lambda^2 -u}{2\Lambda^2}\right)
        .\lab{hyp}
        \eeqa
For $da_D /du$, it is useful to note the analytic continuation formula
        \beq
        _2 F_1 \left(
        \frac{1}{2},\frac{1}{2},1;\frac{\Lambda^2 -u}{2\Lambda^2}\right)=
        \frac{\Gamma(1/2)}{\Gamma(3/4)^2} \ _2 F_1 \left(\frac{1}{4},\frac{1}
        {4},\frac{1}{2};\frac{u^2}{\Lambda^4}\right)
        +\frac{\Gamma(-1/2)}{\Gamma(1/4)^2}\frac{u}{\Lambda^2}
        \ _2 F_1 \left(\frac{3}{4},\frac{3}{4},
        \frac{3}{2};\frac{u^2}{\Lambda^4}\right)        
        .\eeq
Both $a$ and $a_D$ satisfy the following Picard-Fuchs equation
        \beq
        4(\l^4 -u^2)\frac{d^2 \Pi}{du^2}-\Pi =0
        .\eeq
This differential equation can be translated into the 
hypergeometric differential equation of type $_2 F_1 [1/4,1/4,1/2;z]$
        \beq
        z(1-z)\frac{d^2 \Pi}{dz^2}+\left(\frac{1}{2}-\frac{z}{2}\right)
        \frac{d \Pi}{dz}-\frac{\Pi}{16}=0
        ,\eeq
where
        \beq
        z=\frac{u^2}{\l^4}
        .\eeq
        
The solution around $u=\infty$ are given by 
        \beq
        \rho_1 =z^{1/4}\sum_{i=0}^{\infty}\frac{a_i }{z^i},\ \rho_2 =
        \rho_1 \log \frac{1}{z}+z^{1/4}\sum_{i=1}^{\infty}\frac{b_i }{z^i}
        \lab{sol}
        ,\eeq
where 
        \beqa
        a_0 &=&1,\nm\\
        a_1 &=&-\frac{1}{16},\nm\\
        a_2 &=&-\frac{15}{1024},\nm\\
        a_3 &=&-\frac{105}{16384},\nm\\
        a_4 &=&-\frac{15015}{4194304},\nm\\
        a_5 &=&-\frac{153153}{67108864},\nm\\
        a_6 &=&-\frac{6789783}{4294967296},\nm\\
        a_7 &=&-\frac{79676025}{68719476736},\nm\\
        a_8 &=&-\frac{62386327575}{70368744177664},\nm\\
        a_9 &=&-\frac{787916211225}{1125899906842624},\nm\\
        a_{10}&=&-\frac{40814059741455}{72057594037927936}
        \eeqa
and 
        \beqa
        b_1 &=&\frac{1}{8},\nm\\
        b_2 &=&\frac{13}{1024},\nm\\
        b_3 &=&\frac{163}{49152},\nm\\
        b_4 &=&\frac{31183}{25165824},\nm\\
        b_5 &=&\frac{74791}{134217728},\nm\\
        b_6 &=&\frac{7190449}{25769803776},\nm\\
        b_7 &=&\frac{429352037}{2886218022912},\nm\\
        b_8 &=&\frac{161098027539}{1970324836974592},\nm\\
        b_9 &=&\frac{12747539619133}{283726776524341248},\nm\\
        b_{10}&=&\frac{307142061004141}{12970366926827028480}
        .\eeqa
These expansion coefficients can be also obtained from the 
Pochhammer symbols of associated hypergeometric functions, 
but we write explicitly for convenience. 

Direct calculation of the periods fixes the combination of the 
solutions to Picard-Fuchs equations, thus 
        \beq
        a =\frac{\l}{\sqrt{2}}\rho_1 ,\ a_D =i\frac{\l}{\sqrt{2}\pi}
        (-4+6\log 2)\rho_1 -i\frac{\l}{\sqrt{2}\pi}\rho_2 
        .\eeq

The effective coupling constant is then given by 
        \beqa
        \tau_4 &=&\frac{da_D}{du}/\frac{da}{du}\nm\\
        &=&i\frac{2}{\pi}\log\frac{8u}{\l^2}-i\frac{5\l^4}{8\pi u^2}
        -i\frac{269\l^8}{1024\pi u^4}-i\frac{1939\l^{12}}
        {12288\pi u^6}\nm\\
        & &-i\frac{922253\l^{16}}{8388608\pi u^8}-
        i\frac{1394369\l^{20}}{16777216\pi u^{10}}-\cdots
        .\eeqa
        
In order to obtain the prepotential, we need the inversion of $a=a(u)$, 
which is given by
        \beq
        u=2a^2+\frac{\l^4}{16a^2}+\frac{5\l^8}{4096a^6}+
        \frac{9\l^{12}}{131072a^{10}}+\frac{1469\l^{16}}
        {268435456a^{14}}+\cdots
        .\eeq
In this way, we get the prepotential
        \beq
        \F_4 =i\frac{a^2}{\pi}\left[
        \log \left(\frac{a}{\l}\right)^2 +
        4\log 2-3+\sum_{k=1}^{\infty}F_k \left(\frac{\l}{a}
        \right)^{4k}\right]
        ,\eeq
where we have omitted the lower order terms than $O(a)$ and 
the instanton expansion coefficients are  
        \beq
        F_1 =-\frac{1}{64},\ F_2 =-\frac{5}{32768}
        \lab{inexp}
        .\eeq

%%%%%%%%%%%%%%%%%%%%%%%%%%%%%%%%%%%%%%%%%%%%%%%%%%%%%%%%%%%%%%%%%%%%

\begin{center}
\section*{Appendix B. Hypergeometric function}
\end{center}

\renewcommand{\theequation}{B.\arabic{equation}}\setcounter{equation}{0}

Gauss's hypergeometric function $F(\alpha,\beta,\gamma;z)$
often denoted by $ _2 F_1 (\alpha,\beta,\gamma;z)$ satisfies the differential 
equation
        \beq
        z(1-z)\frac{d^2 F}{dz^2} +\left[ \gamma -(\alpha +\beta+1)\right]
        \frac{dF}{dz}-\alpha \beta F =0 
        \lab{Hyp}
        .\eeq
For $|z|<1$, it has a convergent series representation 
        \beqa
       F(\alpha,\beta,\gamma;z)&=&
        \frac{\Gamma(\gamma)}{\Gamma(\alpha)\Gamma(\beta)}
       \sum_{n=0}^{\infty}\frac{\Gamma(\alpha+n)\Gamma(\beta+n)}
        {\Gamma(\gamma+n)}\frac{z^n}{n!}\nm\\
        &=&\sum_{n=0}^{\infty}\frac{(\alpha)_n (\beta)_n }
        {(\gamma)_n}\frac{z^n}{n!}
        ,\eeqa
where $(*)_n =\Gamma(*+n)/\Gamma(*)$ is Pochhammer's symbol.

Using the Euler integral representation
        \beq
        F(\alpha,\beta,\gamma ;z) =
        \frac{\Gamma(\gamma)}{\Gamma(\beta)\Gamma(\gamma -\beta)}
        \int_{0}^{1} t^{\beta -1} (1-t)^{\gamma -\beta -1} 
        (1-zt)^{-\alpha}dt
        ,\eeq
we can establish the formula
        \beqa
        \int_{e_1}^{e_2} \frac{dx}{\sqrt{(x-e_1 )(x-e_2 )(x-e_3 )}}&=&
        \frac{1}{\sqrt{e_3 -e_1}}\int_{0}^{1}
        \frac{dt}{\sqrt{t(1-t)(1-ht)}}\nm\\
        &=&\frac{\pi}{\sqrt{e_3 -e_1 }}F(1/2,1/2,1;h)
        ,\eeqa
where $h=(e_2 -e_1)/(e_3 -e_1)$. Note that the elliptic curve 
        \beq
        y^2 :=(x-e_1 )(x-e_2 )(x-e_3 )
        \eeq
has branching points located at $x=e_i$ $(i=1,2,3)$ and $\infty$.

%%%%%%%%%%%%%%%%%%%%%%%%%%%%%%%%%%%%%%%%%%%%%%%%%%%%%%%%%%%%%%%%%%%%%

\begin{center}
\section*{Appendix C. Schwarzian differential equation}
\end{center}

\renewcommand{\theequation}{C.\arabic{equation}}\setcounter{equation}{0}

We summarize the basics of Schwarzian differential equation in this appendix.

For the two independent solutions denoted by $y_1$ and $y_2$ of the second
order 
differential equation 
        \beq
        \frac{d^2 y}{dx^2}+p(x)\frac{dy}{dx}+q(x)y =0
        ,\eeq
the ratio $z=y_2 /y_1$ is known to satisfy a non-linear differential equation 
called Schwarzian differential equation
        \beq
        \{ z,x\}=-\frac{p^2}{2}-\frac{dp}{dx}+2q
        ,\eeq
where 
        \beq
        \{ z,x\}=\frac{z'''}{z'}-\frac{3}{2}\left(\frac{z''}{z'}\right)^2
        \lab{scw}
        \eeq
is Schwarzian derivative and $' =d/dx$. 

As important properties, it is well-known that (\ref{scw}) satisfies 
        \beq
        \{ y,x\}=-\left(\frac{dy}{dx}\right)^2 \{ x,y\}
        \eeq
and the Cayley's identity 
        \beq
        \{ y,x\}=\left(\frac{dz}{dx}\right)^2 
        \left( \{ y,z\}-\{ x,z\}\right)   
        .\eeq
Furthermore, (\ref{scw}) is invariant under the action of
$SL(2,\mbox{\boldmath$C$})$, i.e., 
        \beq
        \left\{\frac{Ax+B}{Cx+D},x\right\} =\{ z,x\}
        ,\lab{schhu}
        \eeq
where 
        \beq
        \left(\begin{array}{cc}
        A&B\\
        C&D
        \end{array}\right) \in SL(2,\mbox{\boldmath$C$})
        .\eeq
(\ref{schhu}) implies that 
        \beq
        \{ z,x\} =0
        \eeq
has a solution of the form
        \beq
        z=\frac{Ax+B}{Cx+D}
        .\eeq

%%%%%%%%%%%%%%%%%%%%%%%%%%%%%%%%%%%%%%%%%%%%%%%%%%%%%%%%%%%%%%%%%%%%%%%%

\begin{center}
\section*{Appendix D. Polylogarithms}
\end{center}

\renewcommand{\theequation}{D.\arabic{equation}}\setcounter{equation}{0}

In this appendix, we summarize on polylogarithms. For more details, 
see \cite{Lew1,Lew2}.

Polylogarithm $Li_k (x)$ is defined by 
        \beq
        Li_0 (x)=\frac{x}{1-x},\ Li_k (x) =\int_{0}^{x} 
        \frac{Li_{k-1} (t)}{t}dt 
        .\lab{B}
        \eeq
In particular, $Li_2$ is usually called as di-logarithm, 
whereas $Li_3 $ tri-logarithm.
For $|x|<1$, $Li_k $ can be represented by a series
        \beq
        Li_k (x) =\sum_{n=1}^{\infty} \frac{x^n}{n^k}
        .\eeq 

It is easy to show that $Li_k$ $(k>0)$ satisfies a differential 
equation of order $k+1$ \cite{Ram}
        \beq
        \frac{d}{dx}\left[ (1-x)\frac{d}{dx} \left(x\frac{d}{dx}\right)^{k-1}
        \right]y =0.
        \lab{a}
        \eeq
(\ref{a}) is actually equivalent to the generalized hypergeometric 
differential equation 
        \beq
        \left[ x\frac{d}{dx}\prod_{i=1}^{B}\left(x\frac{d}{dx} 
        +\beta_i -1\right)
        -x \prod_{i=1}^{A} \left( x\frac{d}{dx} +\alpha_i \right) \right] 
        \widetilde{y} =0
        \lab{fuc}
        ,\eeq
provided 
        \beq
        A=B+1,\ B=k,\ \alpha_i =1,\ \beta_i =2,\ \widetilde{y} =
        \frac{dy}{dx}
        .\eeq
Note that (\ref{fuc}) is not Fuchsian when $k=1$. 

Let 
        \beq
        (*)_k =\underbrace{(*,*,\cdots,*)}_{k \ \mbox{\scriptsize times}}.
        \eeq
and let the generalized hypergeometric function be $_p F_q $. 
Then we find 
        \beq
        \widetilde{y} =\ _k F_{k-1} \left[ (1)_k ;(2)_{k-1};x\right]
        .\eeq
        \beqa        \frac{dLi_k}{dx} &=& \frac{Li_{k-1}}{x}\nm \\ 
          &=& \ _kF_{k-1} \left[ (1)_k ;(2)_{k-1};x\right]        
      .\eeqa
Thus,   \beq            Li_k (x) =x \ _{k+1} F_{k} \left[
(1)_{k+1};(2)_{k};x\right]            .\lab{soll}             \eeq
The other $k$-independent solutions are given by                \beq       
    1,\ \log x,\cdots,\log^{k-1} x          ,\eeq
where $\log^i x =(\log x)^i$.

Note that 
        \beqa
        Li_k (1) &=& \zeta (k) \nm \\
        &=& \ _{k+1}F_{k} \left[ (1)_{k+1} ;(2)_{k};1\right]
        ,\eeqa
where $\z (k)$ is Riemann's zeta function. 

%%%%%%%%%%%%%%%%%%%%%%%%%%%%%%%%%%%%%%%%%%%%%%%%%%%%%%%%%%%%%%%%%%%%%%

\begin{center}
\section*{Appendix E. Differential equation for prepotential}
\end{center}

\renewcommand{\theequation}{E.\arabic{equation}}\setcounter{equation}{0}

In this appendix, we show that the prepotential can be 
obtained from a differential equation analogous to the 
scaling relation in four dimensions \cite{EY,STY}. 

First, recall that $A_D$ is a differentiation of $\F_5$, i.e., 
$A_D =d\F_5 /dA$. Repeating differentiation over $U$, we obtain 
        \beqa
        \frac{dA_D}{dU}&=&A_{D}'\frac{dA}{dU}=\frac{\F_{5}''}{U '},\nm\\
        \frac{d^2 A_D}{dU^2}&=&\frac{1}{U^{'3}}
        (\F_{5}''' U ' -\F_{5}'' U ''),\nm\\
        \frac{d^3 A_D}{dU^3}&=&\frac{1}{U^{'5}}\left[
        (\F_{5}^{(4)} U ' -\F_{5}'' U ''' )U ' -3 U '' 
        (\F_{5}''' U ' -\F_{5} '' U '' )\right]
        ,\eeqa
where $'=d/dA$. 
Substituting these into the Picard-Fuchs equation for $A_D$ 
and using (\ref{mat2}), we obtain 
        \beq
        \F_{5}^{(4)} +\left[\frac{4R^4 U}{\Del (U )}
        (1-R^4 U^2 )-\frac{1}{U} -3\frac{U ''}{U^{'2}}\right]
        U ' \F_{5} ''' =0
        .\lab{5dif}
        \eeq
In this derivation we do not use Wronskian. This method gives 
a similar equation in the four dimensional theory, but 
the resulting equation is essentially the scaling relation of 
the prepotential. In fact, using (\ref{pff2}) and (\ref{57}) for
the four dimensional theory, one can obtain ($\ \dot{ }
=d/da$) 
        \beq
        a\stackrel{...}{\F}_4 \dot{u} -(
        a\ddot{\F}_4 -\dot{\F}_4 )\ddot{u}=0
        ,\lab{oyo}
        \eeq
which is (twice) integrated to give 
        \beq
        \frac{c_1}{2}u+c_2 =\F_4 -\frac{a}{2}
        \dot{\F}_4 
        ,\lab{ssca}
        \eeq
where $c_i$ are integration constants and for $SU(2)$ pure gauge theory
$c_1 =-i/\pi$ and $c_2 =0$. Then (\ref{ssca}) 
is identified with the scaling relation of the 
prepotential $\F_4$. Note that this derivation of 
the scaling relation is different from those given in \cite{EY,STY}.

(\ref{5dif}) 
can be easily integrated to give 
        \beq
        \F_{5} ''' =c \frac{U U^{'3}}{\Del (U )}
        ,\lab{glo}
        \eeq
where $c$ is an integration constant and is determined by the comparison with 
(\ref{prepo}). The result is $c=-iR^6 /\pi $. 
This is the analogue of the scaling relation of the four dimensional theory. 
We have already calculated 
$U$ as a function of $A$ in section 4.1, so we can determine 
the prepotential by triple integration over $A$. 
In fact, we can check that after suitably choosing 
integration constants appearing in the triple integration 
the prepotential obtained in this way 
agrees to, e.g., the one in the large radius limit (\ref{prepo}). Of course, 
we can extract the information on the prepotential in the small radius 
limit from this relation. Thus the (\ref{glo}) 
is the ``global'' form of prepotential in the five dimensional theory 
in a sense. 

Remark: Triple integration of (\ref{glo}) (with the explicit value of $c$) 
over $A$ gives the prepotential 
itself, but $O(A^2)$-terms are not fixed by this integration (these terms 
include integration constants). The cubic term 
of (\ref{prepo}) or (\ref{prepo2}) is originated from the integration of 
the right hand side of (\ref{glo}). In this sense, (\ref{glo}) 
gives only the cubic term and the infinite sum in (\ref{prepo2}). 

Next, let us consider the relation with the Wronskian of the 
Picard-Fuchs equation. The Wronskian is defined by 
        \beq
        W=\frac{d^2 A}{dU^2}\frac{dA_{D}}{dU}-
        \frac{dA}{dU}\frac{d^2 A_{D}}{dU^2}
        ,\eeq
if the Picard-Fuchs equation (\ref{pf}) is regarded as a second order 
differential equation for $d \Pi/ dU$. 
The Picard-Fuchs equation implies that 
$W$ satisfies the differential equation 
        \beq
        \frac{dW}{dU}=-\left[\frac{4R^4 U}{\Del (U)}
        (1-R^4 U^2)-\frac{1}{U}\right]W
        ,\eeq
which can be integrated to give 
        \beq
        W= \frac{U}{\Del (U)}
        \lab{wrron}
        ,\eeq
up to overall integration constant. This is a very simple 
relation between the Wronskian 
and the moduli. From (\ref{glo}) (with explicit 
value of $c$) and  (\ref{wrron}), we get 
        \beq
        \F_{5} ''' =-i\frac{R^6}{\pi} WU^{'3}
        .\eeq
In the four dimensional theory, integral of Wronskian 
provided the scaling relation. Unfortunatelly, the 
integral of the Wronskian of the five dimansional 
theory does not give such a simple linear relation between 
prepotential and moduli, but the relation between Wronskian 
and the prepotential can be seen from this expression. 

Finally, let us consider the relation between $\F_5$ 
and $\F_4$ by using (\ref{glo}) ($c=-iR^6 /\pi$) 
and (\ref{oyo}). 
Recall that the identification of the 
periods is given by (\ref{Aa}). Using (\ref{Aa}) with the 
differential equation for $u$, (c.f. (\ref{57})) and 
the scaling relation (\ref{ssca}) (with $c_1 =-i/\pi$ and $c_2 =0$), 
we can derive the following relation 
        \beq
        \F_{5} ''' =R^2 U \stackrel{...}{\F}_4
        .\lab{kore}
        \eeq
As is noted in (\ref{expan}), the combination $R^2 U$ 
tends to one for $R\rightarrow 0$. Futhermore, 
in this limit we have $dA \sim da$ (c.f. (\ref{tends})), so we 
can easily integrate (\ref{kore}) to 
give $\F_5 =\F_4 $ (up to terms involving 
integration constants)! This supports 
the result in section 4.4.

%%%%%%%%%%%%%%%%%%%%%%%%%%%%%%%%%%%%%%%%%%%%%%%%%%%%%%%%%%%%%%%%%%%%%%%

\begin{center}

\end{center}

\clearpage

        \begin{figure}[h]
        \begin{center}
        \epsfile{file=fig1.eps,scale=0.8}
        \caption{Relation among singularities}
        \end{center}
        \end{figure}

\end{document}